*Roy Meissner[1], Claudia Ruhland[2], Katja Ihsberner[3]*

# Kompetenzerwerbsförderung durch E-Assessment

Individuelle Kompetenzerfassung am Beispiel des Fachs Mathematik

## 1     Einleitung

In den vergangenen 30 Jahren wurde der Nutzen von Künstlicher Intelligenz im internationalen Raum vorrangig von Vertreter:innen der Informatik und MINT-Fächer in den Schwerpunktthemen (1) Adaptive Systeme und Individualisierung, (2) Assessment und Evaluation, (3) Profiling und Prädiktion sowie (4) Intelligente Tutoring Systeme beforscht [ZA19]. Der vorliegende Beitrag ist im Bereich Assessment verortet und knüpft an Erkenntnisse über die hohe Präzision und Effizienz von E-Assessments bei der Diagnostik von Wissens- und Kompetenzlücken an [ZA19], um Kompetenzen zu erfassen und mithilfe semantischer Webtechnologien zu fördern, wobei Offenheit gegenüber Fachdomänen und differente Zielgruppen gewährleistet wird.

Der Erwerb von Kompetenzen nimmt in Bildungssystemen eine zentrale Rolle ein (vgl. [EU18, pp. 50 ff.]) und kann durch Software-Systeme unterstützt werden. Damit diese, bezogen auf den Erwerb von Kompetenzen, mit hoher Genauigkeit und Effektivität arbeiten können, sind möglichst genaue Daten zu Individuen nötig, aber auch eine didaktische Integration der Anwendungen, wie auch eine maschinenlesbare Aufbereitung von Domänenwissen durch Domänenexpert:innen.

In dieser Veröffentlichung präsentieren wir ein Konzept, wie Micro- und E-Assessments für den Fachbereich Mathematik genutzt werden können, um erworbene und fehlende individuelle Kompetenzen automatisiert zu bestimmen und, davon ausgehend, fehlende bzw. weitere Kompetenzen in einem softwaregestützten Prozess zu erwerben. Die dazu nötigen Modelle sind ein digital aufbereiteter und ausgezeichneter E-Assessment Item-Pool, eine digitale Modellierung der Domäne, die Themen, dafür nötige Kompetenzen als auch einführendes und weiterführendes Material umfasst, sowie ein digitales Individualmodell, das Kompetenzen sicher erfassen kann und auch Aspekte des Kompetenzverlustes integriert.

[1] Universität Leipzig, Institut für Bildungswissenschaften
[2] Freie Universität Berlin, Center für digitale Systeme
[3] Hochschule für Technik, Wirtschaft und Kultur, Fakultät Informatik und Medien

## 2    Kompetenzerfassung durch E-Assessment

### 2.1    E-Assessment

Kompetenzerfassung kann durch verschiedene Methoden operationalisiert werden, im Rahmen des Bologna-Prozesses wird die Formulierung von Learning Outcomes (LOs, *dt*. Lernergebnis) [EU18, pp. 50 ff.] empfohlen. Dabei beziehen wir uns auf die Definition und Erläuterungen aus dem ECTS Users' Guide [EU15, pp. 22 ff.] und gehen davon aus, dass einzelne intendierte LOs zu erwerbendes Wissen und zu erwerbende Kompetenzen (im Folgenden nur noch als Kompetenzen bezeichnet) trennscharf beschreiben sollten. Zur Kategorisierung von Learning Outcomes etablierte sich im deutschsprachigen Raum seit 2001 zunehmend die auf Bloom et al. [BL56] basierende Lernzieltaxonomie von Anderson & Krathwohl [AN01], welche als 4x6-Matrix aus der kognitiven Prozessdimension und der Wissensdimension aufgebaut ist.

Zur Überprüfung von LOs können E-Assessments eingesetzt werden. Diese bestehen aus einer Zusammenstellung einzelner Items, die einer 1:1 Zuordnung (Item zu LO) genügen sollten, welche in den Abschnitten 2.2 und 2.3 für trennscharfe Schlüsse benötigt wird. Außerdem sollten dazu genügend viele Items verschiedener Lerntaxonomiestufen [AN01] pro LO existieren.

### 2.2    Einbettung von E-Assessment Items in ein Domänenmodell

Bei der didaktischen Planung eines Hochschulkurses stehen intendierte LOs im Mittelpunkt und sollen durch aufeinander aufbauende Lerneinheiten erreicht werden. Dieser aufbauende Charakter ermöglicht es, klar zu definieren, welches Vorwissen und welche Kompetenzen zur Bearbeitung eines Themas erforderlich sind, welche Lernan-gebote zur Vermittlung desselben nötig sind und welche Ressourcen die Thematik vertiefen oder erweitern. Die Gesamtheit dieser Informationen bildet ein didaktisch angereichertes Domänenmodell in Form einer Fachlandkarte [MT21] (siehe Abb. 1, oben rechts). Da LOs so inhärent Themen und deren Voraussetzungen semantisch zugeordnet werden können, ist eine transitive Verknüpfung zwischen E-Assessment-Items, und damit -Ergebnissen, und Themen möglich (siehe Abb. 1, rechte Hälfte). Bei entsprechender Ausgestaltung von E- und Micro-Assessments können dadurch nicht nur individuell erworbene Kompetenzen erfasst, sondern auch vermeintlich fehlende Kompetenzen bestimmt werden, indem zusätzlich zum reinem E-Assessment die thematischen Voraussetzungen einbezogen werden, wie in Abschnitt 2.3 dargestellt.

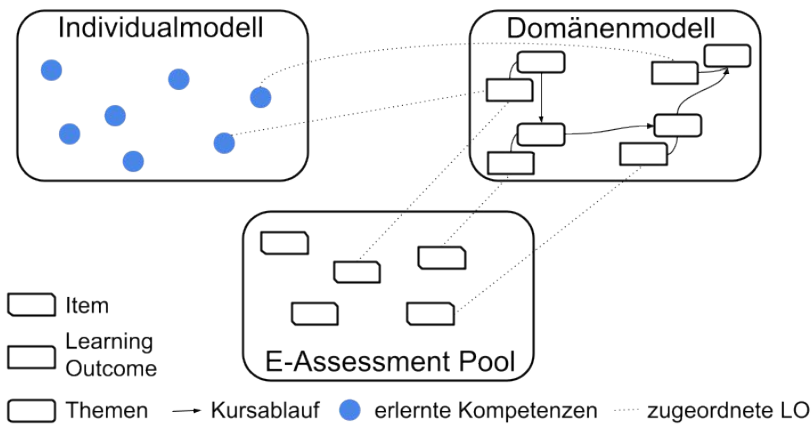

**Abb. 1**: Visualisierung des Lernfortschritts im Zusammenspiel mit angebotenen E-Assessment Items und dem Domänenmodell

## 2.3 Konsequenzen bei der Überlagerung mit dem Individualmodell

Das o.g. Domänenmodell kann mit einem Individualmodell der Lernenden abgeglichen werden, welches deren Wissens- und Kompetenzerwerb individuell aufzeichnet und abbildet. Sind im Domänenmodell determinierte Lernvoraussetzungen nicht in einem Individualmodell repräsentiert, können diese Defizite individuell identifiziert, bspw. durch Micro-Assessments entlang der Taxonomiestufen präzisiert und adäquate Lernressourcen [UL08, pp. 12] entsprechend der zugrundeliegenden Lerntheorie und daraus folgendem didaktischen Instruktionsdesign [GA66, WI00] offeriert werden (siehe nachfolgendes Beispiel). Eine Visualisierung der Schnittmengen und Defizite fördert die metakognitiven Fähigkeiten der Lernenden, indem sie ihre Wissens- und Kompetenzlücken leichter erkennen, durch eine Auswahl passender Lernangebote nach ihren persönlichen Präferenzen gezielt schließen und somit ihren Lernprozess im Sinne des Open Learner Modelling [SE90] effizient und effektiv mitgestalten könnten.

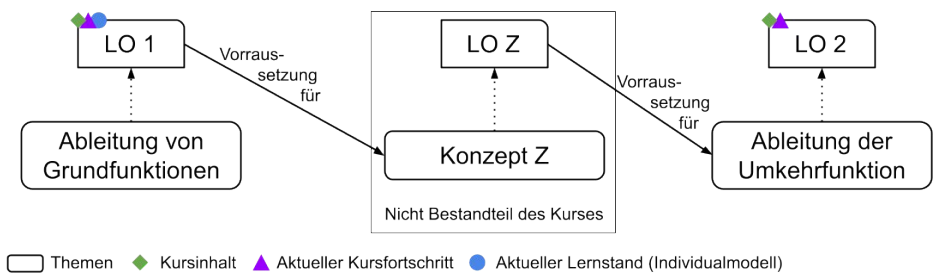

**Abb. 2:** Überlagerung von Individualmodell und Domänenmodellen am Beispiel des Themas „Differenzierbarkeit von Funktionen"

Im Falle einer Abweichung könnte ein Softwaresystem einen konkreten Weg zur Angleichung automatisiert aufzeigen und, je nach didaktischer Modellierung der Domäne, auch Alternativwege anbieten.

Zur Veranschaulichung wird in Abb. 2 eine Lernsituation innerhalb eines Mathematik-Moduls dargestellt, in welcher das Thema *Differenzierbarkeit von*

*Funktionen* behandelt wird. Darin soll das in der Schule erworbene *sichere Beherrschen des Ableitens von Grundfunktionen* (LO 1) aufgefrischt und die sichere *Anwendung der Regel zur Ableitung der Umkehrfunktion* (LO 2) vermittelt werden. Damit Lernende diese Regel leichter lernen und somit nachhaltig in deren Individualmodell verankern können, wird ihnen mithilfe eines Vorschlagsystems eine zusätzliche Lerneinheit (Konzept Z, LO Z) angeboten, beispielsweise zu einem der Konzepte *Kettenregel, Umkehrfunktion* oder *Ableitung von Gleichungen*.

Wird hingegen nur der aktuelle Lernstand über (E-)Assessments regelmäßig erfasst, ist es möglich, dass individuelle Defizite im Vergleich zum Kursfortschritt festgestellt werden. Über die in Abb 2. dargestellte Überlagerung von Domänenmodellen und Individualmodell ist ersichtlich, dass LO 1 erreicht wurde, LO 2 hingegen nicht, wobei LO Z nicht Bestandteil des Kurses ist. Eine Visualisierung der Voraussetzungskette und der Vorschlag entsprechender Lernthemen (Konzept Z in Abb. 2) können beim Erwerb von LO 2 unterstützen.

Neben individuell identifizierten Defiziten ist auch die Erkennung von lernstarken Individuen möglich, die sich ggf. vom Anforderungslevel unterfordert fühlen. Für diese ist es möglich, die subjektiv wahrgenommenen Herausforderungen automatisiert zu erhöhen, beispielsweise indem E-Assessments auf komplexeren Stufen der Lernzieltaxonomie angewendet werden oder indem der zeitliche Rahmen komprimiert wird.

Es ist auch möglich, dass sich das Domänenmodell und Individualmodell nicht überschneiden, wodurch keine Aussage zu fehlenden Kompetenzen getroffen werden kann. Ein Vorteil des didaktisch angereicherten Domänenmodells liegt in seiner Skalierbarkeit, die eine dynamische Erweiterung ermöglicht:

Beispielsweise kann ein modulspezifisches Domänenmodell um Modelle anderer Module ergänzt werden, um so Anküpfungspunkte an das Individualmodell zu finden. Ist solch ein Anküpfungspunkt gefunden, kann durch weitere Micro-Assessments ein individueller Weg zum modulspezifischen Domänenmodell identifiziert und vorgeschlagen werden.

## 3   Zusammenfassung und Ausblick

In diesem Artikel wurde ein Konzept präsentiert, das die automatisierte Bestimmung individuell erworbener und ggf. fehlender Kompetenzen ermöglicht. Dazu werden kompetenzorientierte Individual- und Domänenmodelle abgeglichen, wobei Micro- und E-Assessments zur Kompetenzerfassung und -eingrenzung eingesetzt werden. Die daraus gewonnenen Resultate können in softwaregestützten Prozessen Verwendung finden, bspw. indem durch Lern-Empfehlungssysteme Themen und Materialien mit dem Ziel vorgeschlagen werden, individuellen Kompetenzerwerb zu fördern. Andererseits ist aber auch eine automatisierte Anpassung der subjektiv wahrgenommenen Herausforderung möglich, bspw. um lernstarke Partien zu fordern.

Darüber hinaus stellt die Visualisierung der Fortschritte bzw. bisher erreichter und noch anvisierter Kompetenzen (vgl. Abb. 1 und Abb. 2) eine strukturierende und unterstützende Maßnahme mit Beratungsfunktion für Lernende dar, die im Sinne eines Scaffolding zum Lernerfolg beitragen könnte [Ha09].

Eine prototypische Implementierung des beschriebenen Konzeptes wurde bereits im Rahmen des E-Assessment Literacy Tools Version 2 (EAs.LiT v2) [MR21] und des Fachlandkarten-Tools [MP21] begonnen, deren Fertigstellung den nächsten Schritt im Forschungsdesign darstellen. Weiterhin sollten in zukünftigen Iterationen Konzepte des Kompetenzverlustes (z.B. durch Vergessensprozesse) integriert werden. Letztlich soll eine Kohortenstudie den Einfluss eines softwaregestützten Systems auf den Lernerfolg untersuchen.

**Literatur**